\documentclass[12pt,preprint]{aastex}

\newcommand\be{\begin{equation}}
\newcommand\ee{\end{equation}}

\newcommand{\mcommand}[1]{\ifmmode #1\else $#1$\fi}
\newcommand{\sci}[1]{\mcommand{\times 10^{#1}}}

\newcommand{\unit}[1]{\mcommand{\;{\rm #1}}}
\newcommand\meter{\unit{m}}
\newcommand\second{\unit{s}}
\newcommand\as{\unit{as}}
\newcommand\mas{\unit{mas}}

\newcommand\Mpc{\unit{Mpc}}

\newcommand\mps{\unit{m/s}}
\newcommand\cm{\unit{cm}}
\newcommand\km{\unit{km}}
\newcommand\kmps{\unit{km/s}}

\newcommand\kg{\unit{kg}}

\newcommand\keV{\unit{keV}}

\begin{document}

\slugcomment{Preprint: {\bf CWRU-P02-02}\\
Submitted to {\it The Astrophysical Journal}}

\shorttitle{Improving X-Ray Resolution with {\it XOSS}}
\title{Improving the Resolution of X-Ray Telescopes with Occulting Satellites}
\author{Craig J. Copi\altaffilmark{1} and Glenn D. Starkman\altaffilmark{1}}
\email{cjc5@po.cwru.edu \& gds6@po.cwru.edu}
\altaffiltext{1}{Department of Physics, Case Western Reserve
University, Cleveland, OH 44106-7079}
\authoraddr{10900 Euclid Avenue\\ Cleveland, OH 44106-7079}

\begin{abstract}
  Improving angular resolution is one of X-ray astronomy's big challenges.
  While X-ray interferometry should eventually vastly improve broad-band
  angular resolution, in the near-term, X-ray telescopes will sacrifice
  angular resolution for increased collecting area and energy resolution.
  Natural occultations have long been used to study sources on small
  angular scales, but are limited by short transit times and the rarity of
  transits.  We describe here how one can make use of an {\it X-ray
    Occulting Steerable Satellite\/} ({\it XOSS\/}) to achieve very-high
  resolution of X-ray sources, conventional X-ray telescopes. Similar
  occulting satellites could also be
  deployed in conjunction with future space observatories in other
  wavebands.
\end{abstract} 


\section{Introduction}

One challenge of X-ray astronomy is the relatively low photon fluxes from target sources.
X-ray telescope builders are forced to trade angular resolution 
against increased collecting area.
Thus, while the diffraction limit of a 
$1.2\meter$ aperture X-ray telescope (such as Chandra)
is $0.3\mas$ at $1\keV$,
the $0.5\as$ reality is worse than what is routinely achieved 
at longer wavelengths.
While development proceeds on space-based X-ray interferometers,
near-future X-ray missions such as {\it Constellation X\/} plan to
increase effective area at the price of reduced angular resolution.
{\it MAXIM Pathfinder},
which will afford $100\;\mu$as resolution ,
hopes to launch between 2008 and 2013;  
however, it will have an effective area of only $100\;{\rm cm}^2$.

In the interim, angular resolution can be substantially improved for bright,
relatively stable sources.
The general technique to do this is well-known---eclipse mapping.
When a body such as the moon transits a telescope's field-of-view (FOV),
it occults different sources within the FOV
at different times.  By carefully measuring
the photon count rate as a function of time during the transit,
one can reconstruct the projection of the surface brightness
in the FOV onto the path of the occulter.

Deployment of large occulting satellites has been discussed for the optical
and near infra-red \citep{Adams,Schneider,CS98,CS00}, for both planet
finding and high-resolution astronomy.  However, occulting satellites are
particularly well-suited for use in the X-ray.  In the X-ray, the diffraction
of the photons by the satellite can generally be neglected (see section
\ref{sec:resolve} below).  The optimal size and placement of the satellite
are therefore governed chiefly by one's ability to accurately position the
satellite and station-keep with respect to the line-of-sight to the source.
The achievable resolution is then determined primarily by the telescope
collecting area and by the level to which one can zero the {\it
  XOSS\/}-telescope relative velocity.

We consider an X-ray telescope at the the Earth-Sun system's second
Lagrange point (L2) (such as {\it Constellation X\/}). We discuss the size
and steering of a {\it XOSS\/} in section \ref{sec:buildanduse}.  In
section \ref{sec:resolve} we present
the two-source angular resolution as a function of the {\it
  XOSS}--telescope relative angular velocity and of the photon count rate.
In section \ref{sec:reconstruction} we reconstruct some test sources.
Specific classes of sources are discussed briefly in
section~\ref{sec:results}.  Section~\ref{sec:conclude} contains our
conclusions.

\section{Building and Using a {\it XOSS}}
\label{sec:buildanduse}

\noindent
{\rm \bf  Locating XOSS:}
All current and some future X-ray telescopes use Earth orbit.
Because orbital velocities are so high there, it is 
more difficult to use the approach we describe here; 
we will not discuss these further.
Other X-ray telescopes, such as {\it Constellation X}, will be located at L2
or other low-acceleration environment (e.g.\ Earth-trailing and drift away
orbits).  We focus on these.

We have previously discussed in detail placing a large occulter at L2
\citep{CS00}.  Here we highlight the main points.  Orbits around L2, both
in and out of the ecliptic, have periods $\tau\simeq 6\;{\rm months}$, nearly
independent of their size (for $d \la 10^4\km$).  Consider an X-ray
telescope at L2 and a {\it XOSS} orbiting it at a distance $d$.  The
typical orbital acceleration is $a = 4\pi^2 d / \tau^2$.  Suppose initially
XOSS and the telescope are at relative rest. We require their relative
velocity through a transit to be less than some $v$,
the smaller $v$, the better the expected angular resolution.
For a satellite of size $L$,  the transit duration is $t = L/v$, so
we need $a t < v$.
In terms of the angular velocity, $\mu = v/d$, 
the constraint is 
\be \mu > \!\!  \frac{\sqrt{a L}}{d} 
= 0.26 \sqrt{\left(\frac L{10\meter}\right) \left (\frac{10^6\meter}d \right)} \frac{\mas}{\second}.
\ee
This is the typical relative angular velocity 
assuming no orbital adjustments are made.
Much smaller relative angular velocities can be achieved.
Drifts of $1\meter$ off the line-of-site can  take days,
if the relative-velocity perpendicular to the line-of-sight 
is initially zeroed (using thrusters). 

\ \newline
\noindent
{\rm \bf Building XOSS}
The attenuation length of X-rays in matter is well known~\citep{PDG}.
Except in hydrogen, it is approximately $3\sci{-4} \unit{g/cm^2}$ at
$1\keV$, $10^{-2} \unit{g/cm^2}$ at $10\keV$, and $0.2\unit{g/cm^2}$ at
$100\keV$. 
At a density of $3\unit{g/cm^3}$, these represent thicknesses
of $1$, $30$ and $600$ microns respectively.
An occulter needs to be $\sim3$ attenuation lengths thick,
and so for a $10\meter$ square $\sim1\kg$ to operate at $1\keV$,
$30\kg$ at $10\keV$.
Even at $100\keV$ a $3\meter\times3\meter$ lead film three
attenuation lengths thick would be only $60\kg$.

The ideal size for XOSS depends  on
the telescope aperture, the accuracy with 
which one can position the occulter, and the telescope field-of-view (FOV).  
A typical X-ray aperture, 
which is the minimum occulter size,
is $\sim1\meter$.
Next we estimate how well we can find
{\it XOSS's\/} position relative to the line-of-site.
Let $r$ be the telescope-{\it XOSS\/} separation.
With a diffraction-limited optical telescope of diameter $d$ 
on one spacecraft one can
establish the relative position of the other to 
\be 
\delta x < 1.2 r \frac{\lambda}{d}  
= 0.5 \meter \frac{r}{1000\km} \frac{\lambda/400\unit{nm}}{d/1\meter}
\ee
$1\meter$  accuracy at $1000\unit{km}$ separation
therefore requires a $50\unit{cm}$ finder scope.
With isotropic scattering from the target,
and a reflecting area of $1\meter^{2}$,
sunlight results in a flux of $3\sci{8}/{\rm m}^{2}/{\rm s}$,
at $1000\km$,  suggesting that sufficient light is available to determine
relative positions to $\sim\!1\meter$ or less.

We must also be able to reduce the spacecrafts' relative velocity 
to $\ll\! 1\mps$.
In principle, two position determinations each with error of $\Delta x$,
made a time $t$ apart, determine the velocity within
$\Delta v \simeq \sqrt{2} \Delta x/ t$ (assuming no
error in $t$).  
The ability to reduce $\vert \Delta v\vert$ is then constrained by 
the precision control on the impulse rockets, and by
the maximum time between position determinations.
This time is limited by the orbital accelerations, 
but is sufficiently long ($\gg 10^3\second$) at L2.

Existing or planned X-ray telescopes that could be used with
{\it XOSS\/} have large FOV.  If XOSS covers only part of the FOV, 
the rest will contribute a large background.
This can be mitigated by monitoring each telescope resolution
element.  For {\it Constellation-X\/} with 
$\Delta\theta\!\simeq\!15\as$ this would still require a $70\meter$ XOSS at
$1000\km$ separation,  
though the full resolution element need not be occulted  since some background is acceptable.
Occulters of this scale have been previously considered~\citep{CS00}.  
Still, this provides an important constraint on the size
and distance of {\it XOSS};  alternatively it suggests that an X-ray
telescope with somewhat higher resolution would be a better detector for
{\it XOSS}.

\ \newline
\noindent
{\rm \bf Steering XOSS}
To transit multiple sources it will be necessary
to accelerate the satellite 
both to retarget, and to velocity-match. 
For an areal density of just $1.5\times 10^{-3}\unit{g/cm^2}$
(five attenuation lengths as $1\keV$),
solar radiation pressure causes an acceleration of 
only $4\times10^{-4}\meter/\!\second^2$.
All velocity adjustments will therefore require rockets.

A change $\Delta v$ in the satellite's velocity  is related
by momentum conservation to the mass of propellant ejected
and the velocity of ejection:
\be
\Delta v_{\rm sat}
= \left(\Delta m_{\rm propellant}/m_{\rm sat}\right) v_{\rm ejection}
\ee
If $N$ is the desired number of velocity changes,
then we must keep 
$\Delta m_{\rm propellant}\leq m_{\rm sat}/N$.
Ion engines currently have $v_{\rm ejection}\simeq30 \kmps$, thus
\be
N \leq {30\kmps}/\Delta{ v_{\rm sat}} .
\label{eqn:Ncorrect}
\ee
For satellites separated by $1000\km$ near L2, relative velocities
are only $v_{\rm sat}\!= \!{\cal O}(10^{-4}\kmps)$,
suggesting that propellant supply is not a concern.

\section{Resolution}
\label{sec:resolve}

\begin{figure}
  \plotone{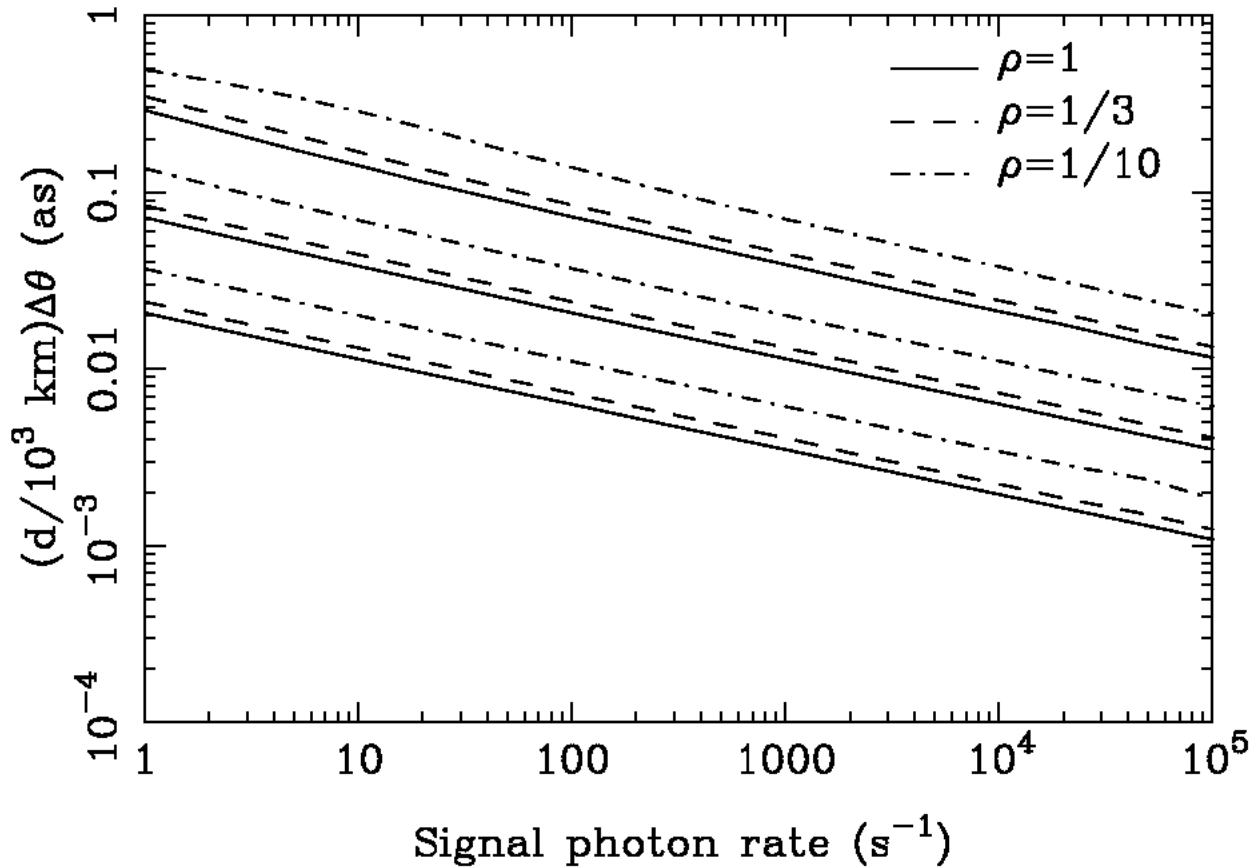}
  \caption{The minimum angular separation of two X-ray sources resolvable
    at the 95\% C.L\@.  Limits are shown for intensity ratios $\rho=1$
    (solid), $1/3$ (dashed), and $1/10$ (dashed-dotted).  The three sets of
    curves are for $\mu_\perp = 10\unit{mas/s}$ (upper),
    $1\unit{mas/s}$ (middle), and $0.1\unit{mas/s}$ (lower).  The
    total photon from both sources detected in (not incident on) the
    unocculted telescope.  A background equal to the flux of the two
    sources is assumed.  Here $d$ is the XOSS-telescope distance.  }
  \label{fig:resolution}
\end{figure}

The angular resolution of the XOSS-telescope system will come from probing
the lightcurve during a transit.  Consider first identifying a binary
source.  For $0.5\keV$ X-rays edge
diffraction from XOSS causes oscillations of order $5\cm$ in the aperture
plane of a telescope $1000\km$ away.  A 1-m aperture telescope averages over
many such oscillations leading to very small deviations from the geometric
shadow, which are further reduced at higher energies.  Determining the
lightcurve therefore reduces to calculating the area of the telescope not
shadowed by XOSS.  An observation consists of a sequence of measurements of
the integrated lightcurve between times $t_{j-1}$ and $t_j$.  We would like
to find the minimum separation of two sources that can be distinguished
from a single source, using such observations.  We therefore calculate the
probability of misidentifying a binary source as a single source.  This
probability depends on $\mu_\perp$, the angular velocity of {\it XOSS\/} as
it transits, and on the photon count rate.  (Throughout we quote photon
rates in the detector, not at the the telescope aperture.  The X-ray
telescope's effective area includes both the geometric collecting area and
the efficiency of the X-ray detector.)  The results for the 95\% confidence
limits as a function of the intensity in a $1.2\meter$ aperture telescope
for $\rho=1$, $1/3$, and $1/10$ and for $\mu_\perp/({\rm mas/s})=10$,
$1$, and $0.1$ are shown in figure~\ref{fig:resolution}.  In producing
figure~\ref{fig:resolution} we assumed a uniform response over the surface
of the telescope.  A more complicated response function may improve
resolution slightly.

This analysis assumed a single transit by a square occulter, yielding
only one projection of the source positions.
One needs multiple projections to resolve a 2-d source.
This could be accomplished by multiple transits, or slits in {\it XOSS\/}
tilted to mimic multiple passes in a single transit.

\section{Image Reconstruction}
\label{sec:reconstruction}

Several techniques have been applied to image reconstruction from 
natural eclipses.  We have focused on the eclipse mapping method 
\citep{BS91}.  EMM is a maximum entropy technique, thus
it can oversmooth the image or introduce spurious sources depending on the
weighting between the entropy and the constraints.
We have implemented an improved EMM algorithm~\citep{BS93}, 
in which we maximize the quality function
\be
  Q = S - {\left[ C (\chi^2) \right]^2}/{\rho} - {\left[ C (R)/\rho
    \right]^2}
\ee
where $S$ is the entropy, the usual definition of
$\chi^2$ is used, $C(x)=(x-x_{\rm aim})/x_{\rm aim}$,
\be
  R = \frac1{\sqrt{M-1}} \sum_{j=1}^{M-1} r_j r_{j+1}
\ee
measures the correlations in the residuals, and $\rho$
controls the weighting between the entropy and the constraints.  We 
anneal in $\rho$ from $100$ to $10^{-4}$,  to allow first the 
entropy to smooth the image then the constraints to sharpen the important
features.  In practice we tend to oversmooth,
leading to a conservative estimate of the attainable resolution.

We modified the EMM algorithm.  
In the geometric limit, sources contribute to changes in the lightcurve 
while they are occulted.  We therefore give more weight to pixels
that are causing the light curve to change.
A minimum pixel intensity is chosen; pixels above threshold count fully 
in the fit while those below the minimum are averaged over.  
Again we anneal from a high value for this intensity threshold to a low value.

\begin{figure}
  \epsscale{0.7}
  \plotone{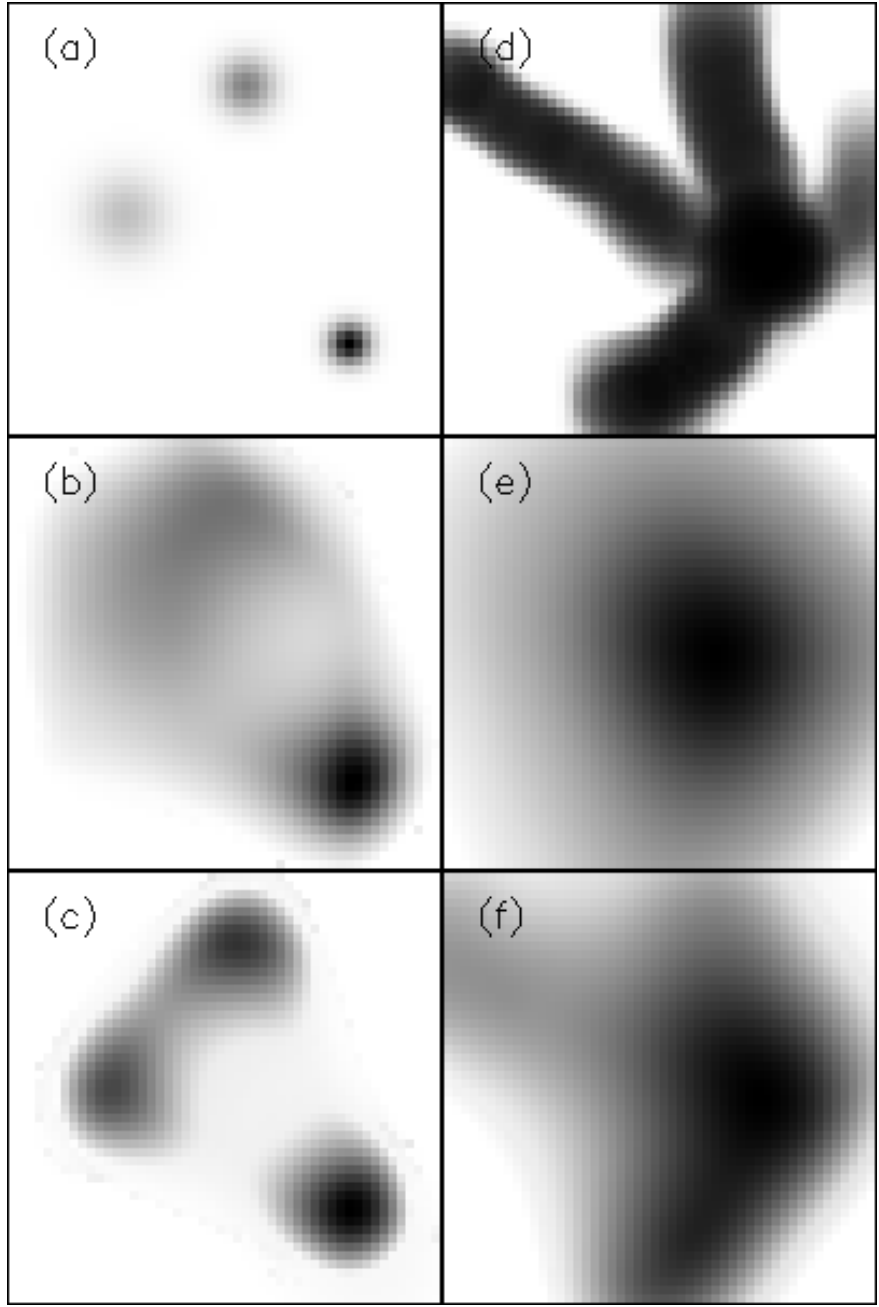}
  \caption{The image reconstruction of (a) Gaussian point sources and (d)
    filaments. Reconstructed images after 4 passes (b and e) and 16 passes
    (c and f) show the reconstruction ability of {\protect\it XOSS}.
  }
  \label{fig:reconstruction}
\end{figure}

To start the reconstruction we could use a low resolution image.  For the
test cases considered here we instead bootstrap by initially performing a
non-negative least squares fit.  Most of the signal is typically placed in
a few very bright pixels---to speed up convergence, we presmooth the image
with a Gaussian.  The smoothing scale must be chosen with some care.
Smoothing on the scale of a few pixels worked well for our test cases.

Two test cases have been considered (Fig.~\ref{fig:reconstruction} panels
(a) and (d) contain the original sources).  The first contains three
Gaussian sources of equal integrated intensity but peak intensities in a
4:2:1 ratio.  The second contains filaments radiating from a bright spot.
The reconstructions are performed for a $10^3 \gamma/{\rm snapshot}$
source.  {\it XOSS\/} moves 1 pixel per snapshot.  The square satellite
provides information along only 1 direction per transit.  Thus we consider
(b,e) 4 and (d,f) 16 passes for each reconstruction.  To extract maximal
information from $N$ passes, each pass the occulter travels at an angle
$\theta = \pi/N$ to the previous pass.

For the Gaussian source the reconstructions are oversmoothed.  By 4 passes
(b) we can resolve at least two sources; with 16 passes (c) all three
sources can be resolved.  For the filamentary source the 4 passes (e)
reconstruct the bright spot well.  With 16 passes (f) the filaments are
resolved.  Even though the reconstruction technique involves Gaussian
smoothing, non-Gaussian structures can be resolved.

\section{Results}
\label{sec:results}

This is an exciting time for X-ray astronomy.  The {\it Chandra} and {\it
  XMM\/} telescopes have been inserted into Earth orbit.  Other major
observatories are being planned for this decade: {\it Astro-E2}, and {\it
  XEUS\/} will enter Earth-orbit, while {\it Constellation X\/} will be
placed at L2.  {\it MAXIM Pathfinder}---the first astronomical X-ray
interferometer---may launch between 2008 and 2013 into a drift-away orbit;
to be followed eventually by {\it MAXIM\/}.  This may seem a remarkable
proliferation, but each mission has its own emphasis.  {\it Chandra\/} is
the only high angular resolution telescope, with
$\Delta\!\theta\!\simeq\!0.5\!\as$, and thus has the lowest effective area.
The other missions (excepting the interferometers) opt instead for large
effective area.

The luminosity of X-ray sources varies greatly.  Black holes in the cores
of nearby galaxies have 
\be 
{\cal L}_{\rm bh} \approx 10^{38\hbox{--}40}\unit{erg/s} =
6.2\sci{46\hbox{--}48} \unit{keV/s}
\ee
in the $0.2\hbox{--}2.4\keV$ energy range.  This leads to a photon rate at
the detector of
\be
\Gamma_{\rm bh} = 6.5\sci{0\hbox{--}2} 
\frac{E_{\keV}}{d_{\rm Mpc}^2 A_{1000}} \second^{-1},
\ee
where $E_{\keV}$ is the X-ray energy in keV, $d_{\rm Mpc}$ is the source
distance in Mpc and ${\cal A}_{1000}$ is the effective area of the
telescope in units of $1000 {\rm cm}^2$.  An AGN, Seyfert galaxy, or the
core of an X-ray cluster can be much more luminous ${\cal L}_{\rm AGN}
\approx 10^{40\hbox{--}44}\unit{erg/s}$; however, since
$d\approx100\Mpc$
\be
\Gamma_{\rm AGN} = 6.5\times 10^{(-2\;{\rm  to }\;+2)}
\frac{10^4 E_{\keV}}{d_{\rm Mpc}^2 A_{1000}} \second^{-1}.
\ee
Galactic microquasars are less luminous
${\cal L}_{\rm microquasar} \approx 10^{39}\unit{erg/s}$,
but, since they are closer
\be
\Gamma_{\rm microquasar} = 6.5\sci{5} 
\frac{10^{-4} E_{\keV}}{d_{\rm Mpc}^2 A_{1000}} \second^{-1}.
\ee

Great improvements in resolution
are readily accessible (figure~\ref{fig:resolution}).  
Sub-milliarcsecond
resolution can be obtained for sources with photon rates $\Gamma\ga
1000\second^{-1}$. For a single {\it Constellation X\/} module, which will
have an effective area of about $3,750\cm^2$, the brightest AGNs, X-ray
cluster cores, and galactic black holes will have $\Gamma\approx
200\second^{-1}$ so we can obtain $\Delta\theta\approx 5\mas$. 

This technique has clearest application to steady sources.
The occulter transit time is 
\be
  t_{\rm transit} = 1000\second \left( \frac{F}{\rm as }\right)
  \left(\frac{\rm mas/s}\mu \right),
\ee
where $F$ is the telescope FOV or resolution, whichever is less.  The
required source-stability timescale is $t_{\rm transit}$ times the number
of passes.  Many extragalactic sources are sufficiently stable.  These
include galaxy clusters, AGNs, quasars, and many X-ray jets; so too are
supernovae.  Galactic sources, however, often vary more quickly.  It may
nevertheless be possible to learn something about them using this
technique.  In particular, optimization of the aperture (e.g. by use of
coded masks) may allow us to significantly reduce both the required transit
time and the required number of transits to obtain a well-resolved 2-d
image.

\section{Conclusions}
\label{sec:conclude}

We have found that a {\it XOSS\/} can lead to tremendous improvements in
angular resolution.  The current trend of increasing the effective area of
X-ray telescopes at the expense of angular resolution until X-ray
interferometry becomes viable, meshes well with the benefits gained by
including a {\it XOSS\/} in the mission.  For the high Earth-orbit X-ray
telescope a moderate improvement in angular resolution over an appreciable
fraction of the sky can be achieved through the use of a {\it XOSS},
however the orbital mechanics are challenging and the scientific pay-off
may be modest.  A technology demonstration mission along these lines may
nevertheless be worthwhile.

{\it XOSS\/} is potentially most interesting for telescopes at L2 and other
low-acceleration environments (e.g.~Earth-trailing or drift-away orbits)
where $1$--$10\mas$ resolution should be achievable for a wide range of
sources.  The ideal telescope would have an angular resolution equal to the
angle subtended by the XOSS.  Further investigations are necessary to
determine whether XOSS is well suited to {\it Constellation X\/}, or
whether a dedicated telescope with better angular resolution would be
merited.  Either way, it seems likely that XOSS could provide a natural
scientific precursor to interferometric missions like {\it MAXIM
  Pathfinder\/}, increasing the angular resolution of X-ray telescopes by
one-to-two orders-of-magnitude at relatively modest incremental expense.

The authors thank A. Chmielewski for support,
A. Babul and M. Dragovan for many useful comments and suggestions,
N. Choudhuri for helpful suggestions on statistical tests, 
P. Gorenstein and W. Zhang for comments on a preliminary version of the 
manuscript and C. Covault and S. Rodney for recent input.
This work was supported by 
a DOE grant to the theoretical particle-astrophysics group at CWRU, 
and a phase I grant from NIAC.

\end{document}